\documentclass[reprint,twocolumn,showpacs,floatfix,prc,aps,10pt]{revtex4-1}
\usepackage{amsmath}
\usepackage{amssymb}
\usepackage{graphicx}
\usepackage{dcolumn}
\usepackage{float}
\usepackage{grffile}
\usepackage{multirow}
\usepackage{txfonts}

\usepackage{hyperref}\hypersetup{colorlinks=true,linkcolor=blue,citecolor=blue,urlcolor=blue,bookmarksnumbered=false,bookmarksopen=false}

\newcommand{\eq}[1]{(\ref{#1})}

\newcommand{\feynp}[1]{#1\kern-0.45em/}
\newcommand{\feynq}[1]{#1\kern-0.46em/}
\newcommand{\s}[1]{\sigma_\textsc{#1}}
\newcommand{\F}{\mathcal{F}}
\newcommand{\GeV}{\;\text{GeV}}
\newcommand{\MeV}{\;\text{MeV}}
\newcommand{\degree}{^\circ}
\newcommand{\ea}{\textit{et al.}}
\newcolumntype{C}[1]{>{\centering\let\newline\\\arraybackslash\hspace{0pt}}m{#1}}
\newcolumntype{L}[1]{>{\left\let\newline\\\arraybackslash\hspace{0pt}}m{#1}}

\renewcommand{\d}{\mathrm{d}}
\renewcommand{\P}{\mathcal{P}}

\graphicspath{{./figures/}}
\DeclareMathAlphabet{\mathcal}{OMS}{cmsy}{m}{n}
\DeclareMathSymbol{\alpha}{0}{letters}{"0B}

\begin{document}

\title{Charged pion electroproduction above the resonance region}

\author{Tom Vrancx}
\email{Tom.Vrancx@UGent.be}
\author{Jan Ryckebusch}
\email{Jan.Ryckebusch@UGent.be}

\affiliation{Department of Physics and Astronomy,\\
Ghent University, Proeftuinstraat 86, B-9000 Gent, Belgium}
\date{\today}

\begin{abstract}
\begin{description}
\item[Background] Above the nucleon resonance region, the $N(e,e'\pi^\pm)N'$ data cannot be explained by conventional hadronic models. For example, the observed magnitude of the transverse cross section is significantly underestimated in a framework with Reggeized background amplitudes.

\item[Purpose] Develop a phenomenological framework for the $N(e,e'\pi^\pm)N'$ reaction at high invariant mass $W$ and deep photon virtuality $Q^2$. 

\item[Method] Building on the work of Kaskulov and Mosel \cite{Kaskulov:2010kf}, a gauged pion-exchange current is introduced with a running cutoff energy for the proton electromagnetic transition form factor. A new transition form factor is proposed. It respects the correct on-shell limit, has a simple physical interpretation and reduces the number of free parameters by one.

\item[Results] A study of the $W$ dependence of the $N(e,e'\pi^\pm)N'$ lends support for the newly proposed transition form factor. In addition, an improved description of the separated and unseparated cross sections at $-t \lesssim 0.5 \GeV^2$ is obtained. The predictions overshoot the measured unseparated cross sections for $-t > 0.5 \GeV^2$. Introducing a strong hadronic form factor in the Reggeized background amplitudes brings the calculations considerably closer to the high $-t$ data.

\item[Conclusions] Hadronic models corrected for resonance/parton duality describe the separated pion electroproduction cross sections above the resonance region reasonably well at low $-t$. In order to validate the applicability of these models at high $-t$, separated cross sections are needed. These are expected to provide a more profound insight into the relevant reaction mechanisms.

\end{description}
\end{abstract}



\maketitle 

\section{Introduction}
\label{sec:intro}
The charged-pion electroproduction reaction $N(e,e'\pi^\pm)N'$ at high energies and deep photon virtuality, is a topic of great theoretical and experimental interest. At high invariant masses, the reaction process is no longer dominated by individual resonances and background contributions prevail in all observables. By increasing the photon virtuality, the electromagnetic charge distribution of the nucleon can be mapped to more and more detail. The hadronic $N(e,e'\pi^\pm)N'$ phenomenology, however, is pushed to its limits in this deep-inelastic regime and is facing complications as partonic degrees of freedom start to overshadow hadronic ones.

The question which mechanisms exactly contribute to the deeply virtual $N(e,e'\pi^\pm)N'$ reaction above the resonance region is a yet unresolved issue which has been around for a few decades \cite{Nachtmann:1976be, Collins:1980dv, Faessler:2007bc}. The hadronic $N(e,e'\pi^\pm)N'$ models fail to reproduce the observed magnitude of the transverse ($\textsc{t}$) and the signs of the interference ($\textsc{tt}$ and $\textsc{lt}$) cross sections in the deep-virtuality, high-energy regime \cite{Blok:2008jy}. In Refs.~\cite{Kaskulov:2008xc, Kaskulov:2009gp}, Kaskulov \ea\ proposed a model which explains the observed features of the transverse cross section. This model is of a hybrid nature: the hadronic background contributions, which dominate at low photon virtuality, are complemented with direct interactions of virtual photons with partons, followed by quark fragmentation into the final nucleon-pion state.

The framework developed in Refs.~\cite{Kaskulov:2008xc, Kaskulov:2009gp} deals with the $N(e,e'\pi^\pm)N'$ reaction at the cross section level. In Ref.~\cite{Kaskulov:2010kf}, Kaskulov and Mosel propose a hadronic model which is able to explain the deep-inelastic $N(e,e'\pi^\pm)N'$ observables at the amplitude level. This model accounts for the residual effects of nucleon resonances in the proton electromagnetic transition form factor. In this approach, nucleon resonances are treated as dual to partons and so the terminology of ``resonance/parton (R/P) contributions'' may be used on occasion. The predictions of the Kaskulov-Mosel (KM) model can be brought in good agreement with the data. However, the electromagnetic form factor for the proton employed in this model is considerably harder than the measured proton Dirac form factor.

In this work, it is shown that the KM model falls short of providing a reasonable description of the data when the employed proton electromagnetic form factor is softened to make it compatible with the Dirac form. An alternative R/P transition form factor is proposed that is both simple and intuitive, and features the accepted cutoff energy scale for the proton electromagnetic form factor. After replacing the transition form factor of the KM model with the newly proposed one, an even better description of the deep-inelastic $N(e,e'\pi^\pm)N'$ data is obtained.

The outline of this work is as follows. In Sec.~\ref{sec:kaskulov_model}, the KM model will be reviewed and discussed. In Sec.~\ref{sec:alt_model}, the R/P transition form factor of the KM model will be addressed and an alternative parametrization will be presented. The comparison of the models with the available deep-inelastic data is the subject of Sec.~\ref{sec:results}. Here, the model predictions are also given for the planned F$\pi$ experiment at the 12 GeV upgrade at Jefferson Lab (JLab) \cite{Huber:2006pr}, and for the recently published CEBAF Large Acceptance Spectrometer (CLAS) $p(e,e'\pi^+)n$ data in the deep pion momentum transfer regime \cite{Park:2012rn}. Finally, the conclusions will be listed in Sec.~\ref{sec:conclusions}.

\section{Kaskulov-Mosel model}
\label{sec:kaskulov_model}

\subsection{Gauged pion-exchange}
\label{subsec:pion_ex}
The main component of the KM model is the gauged pion-exchange current. For the $p(e,e'\pi^+)n$ and the $n(e,e'\pi^-)p$ reaction, these currents read
\begin{align}
J^\mu_{m,m'}(Q^2, s, t) &= i\sqrt{2} g_{\pi N N}\overline{u}_{m'}(p')\gamma_5\Biggl(\F_{\gamma\pi\pi}(Q^2,t,s)\frac{(2k' - q)^\mu}{t - m_\pi^2}\nonumber\\
&\phantom{=} + \F_p(Q^2,s,t)\frac{\feynp{p} + \feynq{q} + m_p}{s - m_p^2}\gamma^\mu\Biggr)u_m(p),\label{eq:J_p}
\end{align}
and
\begin{align}
J^\mu_{m,m'}(Q^2, u, t) &= -i\sqrt{2} g_{\pi N N}\overline{u}_{m'}(p')\Biggl(\F_{\gamma\pi\pi}(Q^2,t,s)\frac{(2k' - q)^\mu}{t - m_\pi^2}\nonumber\\
&\phantom{=} - \F_p(Q^2,u,t)\gamma^\mu\frac{\feynp{p}' - \feynq{q} + m_p}{u - m_p^2}\Biggr)\gamma_5u_m(p),\label{eq:J_n}
\end{align}
in the Lorentz gauge $q\cdot\epsilon = 0$ \cite{Kaskulov:2010kf}. Here, $p$ and $q$ are the four-momenta of the incoming nucleon and virtual photon, $k'$ and $p'$ are the four-momenta of the outgoing pion and nucleon, and $\epsilon$ is the photon polarization four-vector. These kinematics are defined in the laboratory frame. The photon virtuality is defined as $Q^2 = -q^2$, and the Mandelstam variables are given by $t = (k' - q)^2$, $s = W^2 = (p + q)^2$, and $u = (p' - q)^2$. The spin indices of the incoming and outgoing nucleon are denoted by $m$ and $m'$. Further, $g_{\pi NN}$ is the pion-nucleon coupling constant and is fixed at $g_{\pi NN} = 13.4$ in the KM model. The $\F_{\gamma\pi\pi}(Q^2, t, s)$ and $\F_p(Q^2, s, t)$ (where ``$s$'' is interchangeably used for $s$ and $u$) represent the transition form factors of the intermediate pion and proton. The pion-exchange currents, as defined in Eqs.~\eq{eq:J_p} and \eq{eq:J_n}, are gauge-invariant: $q\cdot J_{m,m'} = 0$.

The separated cross sections $\d\s{\{L,T,TT,LT\}}/\d t$ are defined as in Eqs.\ (A1--A4) of Ref.~\cite{Kaskulov:2010kf}. Throughout this work, 
``$\s{\{U,L,T,TT,LT\}}$'' will often be used as a shorthand notation for $\d\s{\{U,L,T,TT,LT\}}/\d t$ or $\d\s{\{U,L,T,TT,LT\}}/\d \Omega_\pi$. Here, $\s{u}$ denotes the unseparated cross section and is given by
\begin{align}
\s{u} = \s{t} + \varepsilon\s{l},
\end{align}
with $\varepsilon$ the ratio of longitudinal to transverse polarization of the virtual photon (Eq.\ (8) of Ref.~\cite{Kaskulov:2010kf}).

The pion transition form factor $\F_{\gamma\pi\pi}(Q^2, t, s)$ is defined as
\begin{align}
\F_{\gamma\pi\pi}(Q^2, t, s) = F_{\gamma\pi\pi}(Q^2)\P_\pi(t, s)(t - m_\pi^2),
\end{align}
with $\P_\pi(t, s)$ the degenerate $\pi(140)/b_1(1235)$-Regge propagator:
\begin{align}
\P_\pi(t, s) = -\alpha_\pi'\varphi_\pi(t)\Gamma(-\alpha_\pi(t))(\alpha_\pi's)^{\alpha_\pi(t)}.
\end{align}
Here,
\begin{align}
\alpha_\pi(t) = \alpha_\pi'(t - m_\pi^2),
\end{align}
is the corresponding Regge trajectory with $\alpha_\pi' = 0.74 \GeV^{-2}$. The Regge phase $\varphi_\pi(t)$ is given by
\begin{align}
\varphi_\pi(t) =
\begin{cases}
e^{-i\pi\alpha_\pi(t)} &p(e,e'\pi^+)n, \\
1 &n(e,e'\pi^-)p.
\end{cases}
\end{align}
For the pion transition form factor $F_{\gamma\pi\pi}(Q^2)$ a monopole parametrization is used:
\begin{align}
F_{\gamma\pi\pi}(Q^2) = \Biggl(1 + \frac{Q^2}{\Lambda_{\gamma\pi\pi}^2}\Biggr)^{-1},
\end{align}
with $\Lambda_{\gamma\pi\pi}$ the pion cutoff energy. In order for the currents $J^\mu_{m,m'}$ to remain gauge-invariant, the transition form factors must coincide at the real-photon point:
\begin{align}
\F_{\gamma\pi\pi}(Q^2 = 0, t, s) = \F_p(Q^2 = 0, s,t),
\end{align}
which implies that the proton transition form factor must be proportional to the pion-Regge propagator:
\begin{align}
\F_p(Q^2, s,t) = F_p(Q^2, s)\P_\pi(t, s)(t - m_\pi^2).\label{eq:F_p-gauge}
\end{align}
In the KM model an additional anti-shrinkage effect in $\F_p(Q^2, s,t)$ is taken into account. More specifically, the slope of the $\pi(140)/b_1(1235)$-Regge trajectory in the Regge propagator of Eq.~\eq{eq:F_p-gauge} is altered as follows:
\begin{align}
\alpha_\pi' \to \alpha_\pi'(Q^2, s) = \frac{\alpha_\pi'}{1 + a\frac{Q^2}{s}},\label{eq:anti-shrinkage}
\end{align}
with $a > 0$. The incorporation of this anti-shrinkage effect does not violate gauge invariance as
\begin{align}
\alpha_\pi'(Q^2 = 0, s) = \alpha_\pi'.
\end{align}

\subsection{Nucleon resonances}
\label{subsec:resonances}
In Ref.~\cite{Kaskulov:2010kf}, it is argued that assigning the proton Dirac form factor to $F_p(Q^2, s)$ might be too naive, as the intermediate proton becomes highly off-shell. A transition form factor which accounts for the effects of the virtual proton fluctuating into resonances was proposed:
\begin{align}
F_p(Q^2, s) = \dfrac{\displaystyle\lim_{\varepsilon \to 0^+}\int_{m_p^2}^\infty\d s_i\dfrac{s_i^{-\beta}}{s - s_i + i\varepsilon}\Biggl(1 + \xi \frac{Q^2}{s_i}\Biggr)^{-2}}{\displaystyle\lim_{\varepsilon \to 0^+}\int_{m_p^2}^\infty\d s_i\dfrac{s_i^{-\beta}}{s - s_i + i\varepsilon}}.\label{eq:KM_RPFF}
\end{align}
The above integral is the continuation of an infinite sum running over all the proton resonances with squared mass $s_i$. The factor $s_i^{-\beta}$, with $\beta \ge 1$ a fit parameter, accounts for the electromagnetic and the strong proton-resonance coupling strengths, and the density of resonance states. The factor
\begin{align}
F_{r_i}(Q^2, s_i) = \Biggl(1 + \xi \frac{Q^2}{s_i}\Biggr)^{-2},\label{eq:FF_resonance}
\end{align}
is a dipole parametrization for the electromagnetic form factor of the resonance $r_i$. Here, $\xi$ is a common average cutoff parameter. In the $s$-channel the singularity at $s_i = s + i\varepsilon$ generates an imaginary part for the proton transition form factor, which is absent in the $u$-channel where the singularity resides in the unphysical region.

\subsection{Additional Regge exchanges and model parameters}
The KM model features two additional Regge exchanges: the vector $\rho(770)/a_1(1320)$ and the axial-vector $a_1(1260)$ trajectories. The exchange currents and accompanying parameters for these amplitudes are listed in Sec.~V and Table I of Ref.~\cite{Kaskulov:2010kf}.

In the KM model, the parameters $(a, \beta, \xi)$ introduced in Eqs.~\eq{eq:anti-shrinkage} and \eq{eq:KM_RPFF} adopt the values
\begin{align}
a = 2.4,\qquad \beta = 3,\qquad \xi = 0.4,
\end{align}
and the following prescription for the pion cutoff energy $\Lambda_{\gamma\pi\pi}$ is employed (see last paragraph of Sec.~VI of Ref.~\cite{Kaskulov:2010kf}):
\begin{align}
\Lambda_{\gamma\pi\pi} \simeq  
\begin{cases}
775 \MeV &Q^2 < 0.4 \GeV^2,\\
630 \MeV &0.6 < Q^2 < 1.5 \GeV^2,\\
680 \MeV &\text{``deep $(Q^2, W)$ region''}.
\end{cases}
\label{eq:pion_cutoff-KM}
\end{align}

\begin{figure}[!b]
\centering
\includegraphics[scale=1]{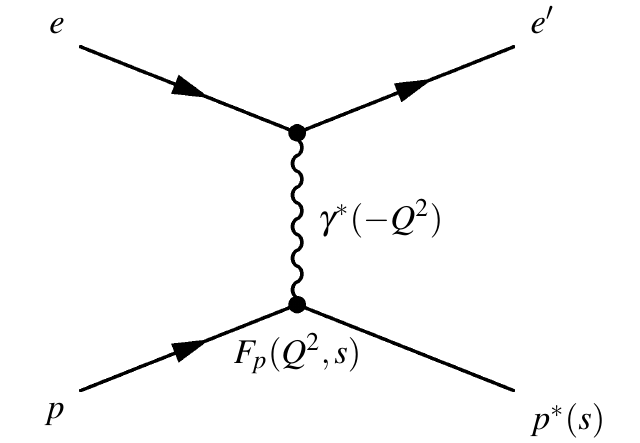}
\caption{Feynman diagram of a virtual photon exciting an incoming proton to the $s$-channel.}
\label{fig:EM_vertex}
\end{figure}

\section{Transition form factor}
\label{sec:alt_model}

\subsection{On-shell limit of the KM transition form factor}
\label{subsec:on-shell_limit}
The transition form factor $F_p(Q^2, s)$ essentially quantifies the R/P contributions to the electromagnetic coupling strength of the intermediate proton in the gauged pion-exchange diagram. The tree-level Feynman diagram of this process is depicted in Fig.~\ref{fig:EM_vertex} for the $s$-channel. The R/P transition form factor of Eq.~\eq{eq:KM_RPFF} was developed independently from the $N(e,e'\pi^\pm)N'$ reaction and could find application in any reaction which has a virtual photon coupling between an on-shell and an off-shell proton.

\begin{figure}[!t]
\begin{center}
\footnotesize
\input{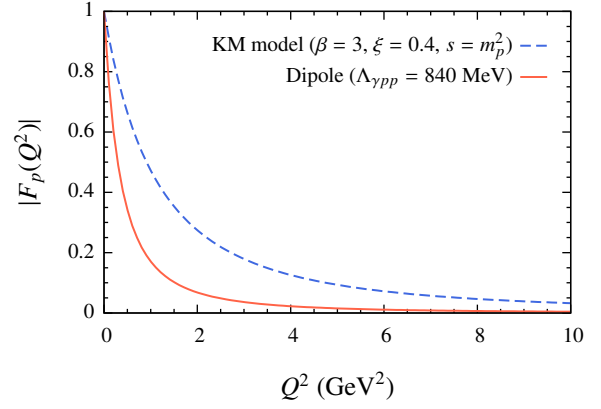}
\\ $ $ \\ $ $ \\
\end{center}
\caption{(Color online) The modulus of the R/P transition form factor of the KM model in the limit $s \to m_p^2$ (dashed lines), and of the proton dipole form factor (full line).}
\label{fig:RP_formfactor}
\end{figure}

In the limit $s \to m_p^2$, the proton remains on its mass shell and the process depicted in Fig.~\ref{fig:EM_vertex} describes elastic electron-proton scattering. As a consequence, one expects that $F_p(Q^2, s)$ reduces to the Dirac form factor $F_p^{\text{Dirac}}(Q^2)$:
\begin{align}
\lim_{s \to m_p^2}F_p(Q^2, s) = F_p^{\text{Dirac}}(Q^2),\label{eq:consistency}
\end{align}
which can be approximated by a dipole form factor (especially at low $Q^2$):
\begin{align}
F_p^{\text{Dirac}}(Q^2) \simeq \Biggl(1 + \frac{Q^2}{\Lambda_{\gamma pp}^2}\Biggr)^{-2},
\end{align}
with $\Lambda_{\gamma pp} = 840\MeV$. As becomes apparent from Fig.~\ref{fig:RP_formfactor}, the transition form factor of Eq.~\eq{eq:KM_RPFF} does not satisfy this constraint and corresponds with a form factor which is considerably harder than $F_p^{\text{Dirac}}(Q^2)$. In that respect, it should be mentioned that the KM framework was developed based on duality arguments, which are only meaningful at higher invariant masses.

The transition form factor of Eq.~\eq{eq:KM_RPFF} is composed of a smooth, infinite distribution of resonance form factors $F_{r_i}(Q^2, s_i)$, which are defined in Eq.~\eq{eq:FF_resonance}. The ground state ``$r_0$'' of those resonances is the proton ($s_0 = m_p^2$). This implies that the form factor for the proton, adopted in the KM model, reads
\begin{align}
F_{r_0}(Q^2, s_0) = \Biggl(1 + \xi \frac{Q^2}{m_p^2}\Biggr)^{-2}.
\end{align}
It can be shown that $\lim_{s\to m_p^2}F_p(Q^2, s) = F_{r_0}(Q^2, s_0)$, as expected intuitively. Hence, the proton cutoff energy used in the KM model amounts to
\begin{align}
\Lambda_{\gamma pp} \to \frac{m_p}{\sqrt{\xi}} \simeq 1484\MeV,
\end{align}
which is considerably larger than $840$ MeV. In order to impose the $s \to m_p^2$ limit of Eq.~\eq{eq:consistency} to the KM transition form factor a value
\begin{align}
\xi = \frac{m_p^2}{\Lambda_{\gamma pp}^2} \simeq 1.248, \label{eq:xi_constraint}
\end{align}
is required. The KM model with $\xi = 1.248$ and $\beta$ the only remaining free parameter, will be dubbed the ``constrained Kaskulov-Mosel'' or ``cKM'' model.

\begin{figure*}[!t]
\begin{center}
\footnotesize
\input{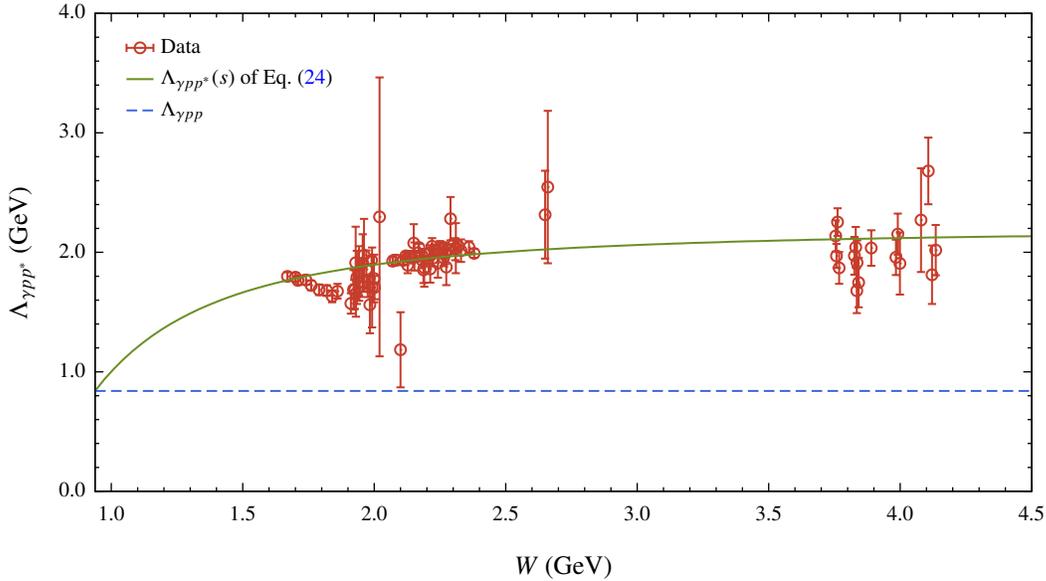}
\\ $ $ \\ $ $ \\
\end{center}
\caption{(Color online) The fitted proton cutoff energy as a function of $W$. The ``data'' are the cutoff values extracted from fitting the VR model (with $g_{\pi N N} = 13.0$, $\Lambda_{\gamma \pi \pi} = 655\MeV$, and $a = 2.4$) to sets of experimental observables (see text). The full line corresponds with the $s$-dependent proton cutoff energy of Eq.~\eq{eq:cutoff_s} for $\Lambda_\infty = 2194\MeV$. The dashed line shows $\Lambda_{\gamma pp} = 840\MeV$.}
\label{fig:proton_cutoff}
\end{figure*}

\subsection{Alternate transition form factor}
\label{subsec:consistent_FF}
Apart from not respecting the constraint \eq{eq:consistency}, the KM prescription \eq{eq:KM_RPFF} for the proton transition form factor has a complex functional dependence on its variables and parameters. Consider the following phenomenological $s$-dependent transition form factor
\begin{align}
F_p(Q^2, s) = \Biggl(1 + \frac{Q^2}{\Lambda_{\gamma p p^*}^2(s)}\Biggr)^{-2}, \label{eq:FF_asymptotic}
\end{align}
which is of the dipole form and has an $s$-dependent cutoff energy $\Lambda_{\gamma p p^*}(s)$. On-shell consistency requires that
\begin{align}
\Lambda_{\gamma p p^*}(m_p^2) = \Lambda_{\gamma pp}.
\end{align}
From the observed magnitude of $\s{t}$ at high energies, it is conceived that an $s$-channel cutoff energy much larger than $\Lambda_{\gamma pp}$ is required. For the transition form factor \eq{eq:FF_asymptotic}, this implies that $\Lambda_{\gamma p p^*}(s)$ should grow with $s$. Assuming that $\Lambda_{\gamma p p^*}(s \to \infty)$ approaches a constant value $\Lambda_\infty$, the lowest-order (with respect to $s^{-1}$) ansatz for the cutoff energy reads
\begin{align}
\Lambda_{\gamma p p^*}(s) = \Lambda_{\gamma pp} + (\Lambda_\infty - \Lambda_{\gamma pp})\Biggl(1 - \frac{m_p^2}{s}\Biggr),\label{eq:cutoff_s}
\end{align}
for $s \ge m_p^2$. For $u \le m_p^2$, the symmetrization of the above expression (for $s \ge m_p^2$) about $m_p^2$ will be employed:
\begin{align}
\Lambda_{\gamma p p^*}(u) = \Lambda_{\gamma pp} + (\Lambda_\infty - \Lambda_{\gamma pp})\Biggl(1 - \frac{m_p^2}{2m_p^2 - u}\Biggr).
\end{align}
The form factor of Eq.~\eq{eq:FF_asymptotic} has a very intuitive $(Q^2,s)$ dependence: an exponential charge distribution is assigned to the proton and the charge radius asymptotically decreases with increasing virtuality.

\subsection{Pion coupling strengths}
\label{subsec:pion_coupling}

In the KM model, a $(Q^2,s)$-dependent parametrization \eq{eq:pion_cutoff-KM} for the pion cutoff energy $\Lambda_{\gamma\pi\pi}$ is employed. This parametrization is discontinuous with respect to $Q^2$ and $s$. In the new model, a constant pion cutoff energy will be adopted which is the average of the upper and intermediate values used in the KM model (respectively $680\MeV$ and $630\MeV$):
\begin{align}
\Lambda_{\gamma\pi\pi} = 655\MeV. \label{eq:pion_cutoff}
\end{align}
In analogy with the off-shell proton case, a $t$-dependent pion cutoff energy could be adopted which amounts to the vector-meson dominance value of $\Lambda_{\gamma\pi\pi} = m_{\rho(770)} \simeq 775.5\MeV$ for $t = m_\pi^2$. However, as the available data only covers a small range of $-t$ values ($-t \lesssim 0.5\GeV^2$), a constant pion cutoff energy can be used. Note that as the $t$-channel pion exchange is replaced by the exchange of a pion-Regge trajectory, the relation to the on-shell pion form factor might be lost and $\Lambda_{\gamma\pi\pi}$ should be interpreted as an effective transition cutoff energy.

The `exact' value of the pion-nucleon coupling $g_{\pi NN}$ is a matter of debate in the literature. Reported values vary from $g_{\pi NN} \simeq 13.0-13.5$ \cite{Gibbs:1998, Rentmeester:1999, Ericson:2000, Pavan:2000}. As mentioned, the value $g_{\pi NN} = 13.4$ is used in the KM model. However, a better agreement with the available $N(e,e'\pi^\pm)N'$ data can be obtained with $g_{\pi NN} = 13.0$ and this value will be used in the new model. The new model will now be dubbed the ``Vrancx-Ryckebusch'' or ``VR'' model.

\begin{table*}[!htbp]
\caption{Comparison of the $\chi^2_\textsc{ndf}$ values for the available $p(e,e'\pi^+)n$ and $n(e,e'\pi^-)p$ DIS data between the KM, the cKM, and the VR model.}
\label{tab:chi2}
\renewcommand{\arraystretch}{1.15}
\begin{tabular*}{\textwidth}{@{\extracolsep{\fill}} llcccccc}
\hline
\noalign{\smallskip}\hline
& \multirow{2}{*}{Observable(s)} & \multirow{2}{*}{$W$ (GeV)} & \multirow{2}{*}{$Q^2$ (GeV$^2$)} &  & \multicolumn{3}{c}{$\chi^2_\textsc{ndf}$} \\ \cline{6-8}
& & &  &  & KM & cKM & VR \\
\hline
CEA & $\s{lt}$ & 2.02 \;--\; 2.31 & 0.354 \;--\; 0.426 & \cite{Brown:1973wr} & 0.71 & 2.79 & 0.52 \\
& $\s{tt}, \s{lt}$ & 2.15 & 0.176 \;--\; 0.188 & \cite{Brown:1973wr} & 0.78 & 1.35 & 0.99 \\
Cornell & $\s{u}, \s{tt}, \s{lt}$ & 2.66 & 1.20 & \cite{Bebek:1974iz} & 0.96 & 1.20 & 1.56 \\
& $\s{l}, \s{t}$ & 2.15 \;--\; 2.65 & 1.19 \;--\; 3.32 & \cite{Bebek:1976qm} & 2.31 & 2.06 & 1.87 \\
DESY & $\s{l}, \s{t}, \s{tt}, \s{lt}$ & 2.10 & 0.35 & \cite{Ackermann:1977rp} & 1.69 & 0.95 & 3.27 \\
& \multirow{2}{*}{$\s{u}, \s{tt}, \s{lt}$}\; $(p)$ & \multirow{2}{*}{2.19} & \multirow{2}{*}{0.70 \;--\; 1.35} & \multirow{2}{*}{\cite{Brauel:1979zk}} & 3.18 & 6.61 & 2.77 \\
& \phantom{$\s{u}, \s{tt}, \s{lt}$}\; $(n)$ &  &  & & 2.22 & 1.55 & 1.37 \\
& $\s{l}, \s{t}, \s{tt}, \s{lt}$ & 2.19 & 0.70 & \cite{Brauel:1979zk} & 2.59 & 3.06 & 3.24 \\
& $\s{u}$ & 3.768 \;--\; 4.121 & 1.37 \;--\; 5.44 & \cite{Airapetian:2007aa} & 4.37 & 74.5 & 3.59 \\
JLab & $\s{l}, \s{t}, \s{tt}, \s{lt}$ & 1.911 \;--\; 2.001 & 0.526 \;--\; 1.702 & \cite{Tadevosyan:2007yd} & 8.11 & 7.95 & 6.03 \\
& $\s{l}, \s{t}, \s{tt}, \s{lt}$ & 2.153 \;--\; 2.308 & 1.416 \;--\; 2.703 & \cite{Blok:2008jy} & 4.84 & 31.8 & 3.96 \\
& $\s{l}, \s{t}, \s{tt}, \s{lt}$ & 2.21 \;--\; 2.22 & 2.15 \;--\; 3.91 & \cite{Horn:2007ug} & 3.37 & 14.1 & 3.29 \\
& $\s{u}$ & 1.70 \;--\; 2.38 & 0.92 \;--\; 4.98 & \cite{Qian:2009aa} & 4.48 & 5.10 & 3.70 \\
& $A_\textsc{lu}^{\sin\phi_\pi}$ & 2.0 & 1.5 & \cite{Avakian:2004dt} & 0.43 & 9.08 & 0.96 \\
\hline
& &  & \multicolumn{2}{r}{Total} & 3.97 & 11.0 & 3.33 \\
\hline
\noalign{\smallskip}\hline
\end{tabular*}
\end{table*}

\section{Results}
\label{sec:results}

\subsection{Proton cutoff energy}
\label{subsec:proton_cutoff_energy}
The VR model features only one parameter ($\Lambda_\infty$) for the proton transition form factor, instead of two ($\beta$ and $\xi$) for the KM model. Before determining the value of the asymptotic proton cutoff energy $\Lambda_\infty$, the experimental energy dependence of the proton cutoff energy will be investigated. To that end, the cutoff energy $\Lambda_{\gamma p p^*}$ of the dipole transition form factor \eq{eq:FF_asymptotic} is fitted to each set of observables ($\s{u}$, $\s{l}$, $\s{t}$, $\s{tt}$, and/or $\s{lt}$) at a fixed invariant mass $W$ and varying $Q^2$ and/or $t$ values. The data employed are from CEA \cite{Brown:1973wr}, Cornell \cite{Bebek:1974iz, Bebek:1976qm}, DESY \cite{Ackermann:1977rp, Brauel:1979zk, Airapetian:2007aa}, and JLab\cite{Tadevosyan:2007yd, Blok:2008jy, Horn:2007ug, Qian:2009aa}.

In Fig.~\ref{fig:proton_cutoff}, the fitted proton cutoff energies $\Lambda_{\gamma p p^*}$ are shown as a function of $W$. There is clear evidence that the high-energy data require a proton cutoff energy much larger than $\Lambda_{\gamma p p} = 840\MeV$. The fitted cutoff energies rise slowly with the energy in the region $W \simeq 1.7 - 2.4\GeV$ and tend to reach a certain asymptotic value. The `experimental' cutoff energies can be well described by the function $\Lambda_{\gamma p p^*}(s)$ of Eq.~\eq{eq:cutoff_s}. Optimizing this function against the extracted cutoff energies yields
\begin{align}
\Lambda_\infty = 2194 \pm 13\MeV,\label{eq:cutoff_infty}
\end{align}
with $\chi^2_\textsc{ndf} = 1.97$ for 84 degrees of freedom. This value is about 2.6 times larger than the on-shell proton cutoff energy. In the VR model, $\Lambda_\infty$ is fixed to the value of Eq.~\eq{eq:cutoff_infty}.

\subsection{Low \texorpdfstring{$-t$}{-t} regime}
\label{subsec:low_t}
At low momentum transfer in the $t$-channel, the final pion is produced at low scattering angles. In this regime, the exchanged pion-Regge trajectory is close to its first materialization (the pion), which results in a dominant longitudinal and a small transverse contribution to the differential cross section. Experimentally, however, it is observed that in the deep-inelastic scattering (DIS) regime the differential cross section receives a sizable contribution from the transverse component. In the current framework, this transverse strength is provided by contributions from resonances/partons to the $s$- or $u$-channel gauge-fixing terms of Eqs.~\eq{eq:J_p} and Eqs.~\eq{eq:J_n}.

In Table \ref{tab:chi2}, the $\chi^2_\textsc{ndf}$ values for the available $p(e,e'\pi^+)n$ and $n(e,e'\pi^-)p$ DIS data are listed for the KM, the cKM, and the VR model. For the cKM model, the value $\beta = 1$ was found to be in best agreement with the data. The cKM model provides a far worse description of the data ($\chi^2_\textsc{ndf} = 11.0$) than the KM model ($\chi^2_\textsc{ndf} = 3.97$). Hence, the KM framework cannot provide a fair description of the data once the correct on-shell limit of the proton electromagnetic transition form factor is imposed.

\begin{figure*}[!t]
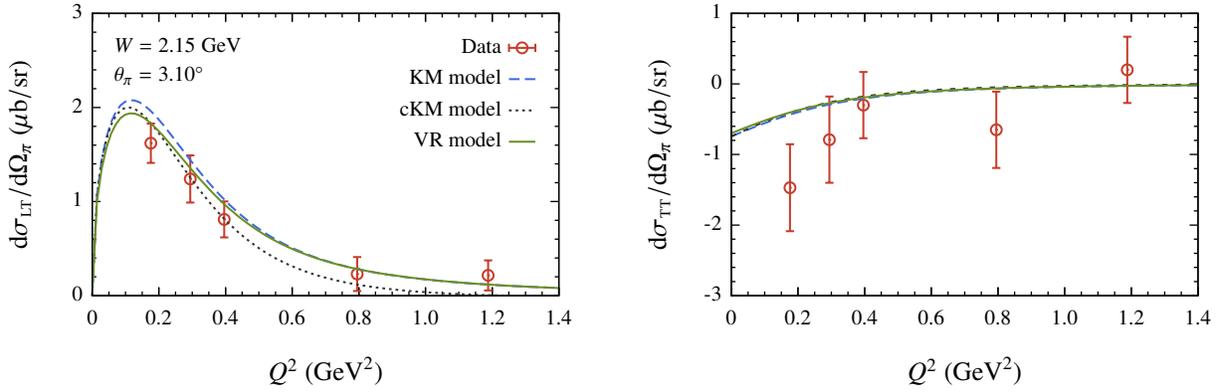

\begin{center}
\footnotesize
\qquad\quad
\input{CEA-diff_LT-fig}\qquad\qquad\qquad\qquad
\input{CEA-diff_TT-fig}
\\ $ $ \\ $ $ \\
\end{center}
\caption{(Color online) The $Q^2$ dependence of the interference cross sections $\d\s{lt}/\d\Omega_\pi$ and $\d\s{tt}/\d\Omega_\pi$ of the $p(e,e'\pi^+)n$ reaction at forward scattering ($\theta_\pi = 3.10\degree$) and $W = 2.15\GeV$. The dashed curves, the dotted (dark gray), and the full curves are the predictions of the KM, the cKM, and the VR model. The data are from Ref.~\cite{Brown:1973wr} (CEA).}
\label{fig:CEA-diff_TT-LT}
\end{figure*}

\begin{figure*}[!t]
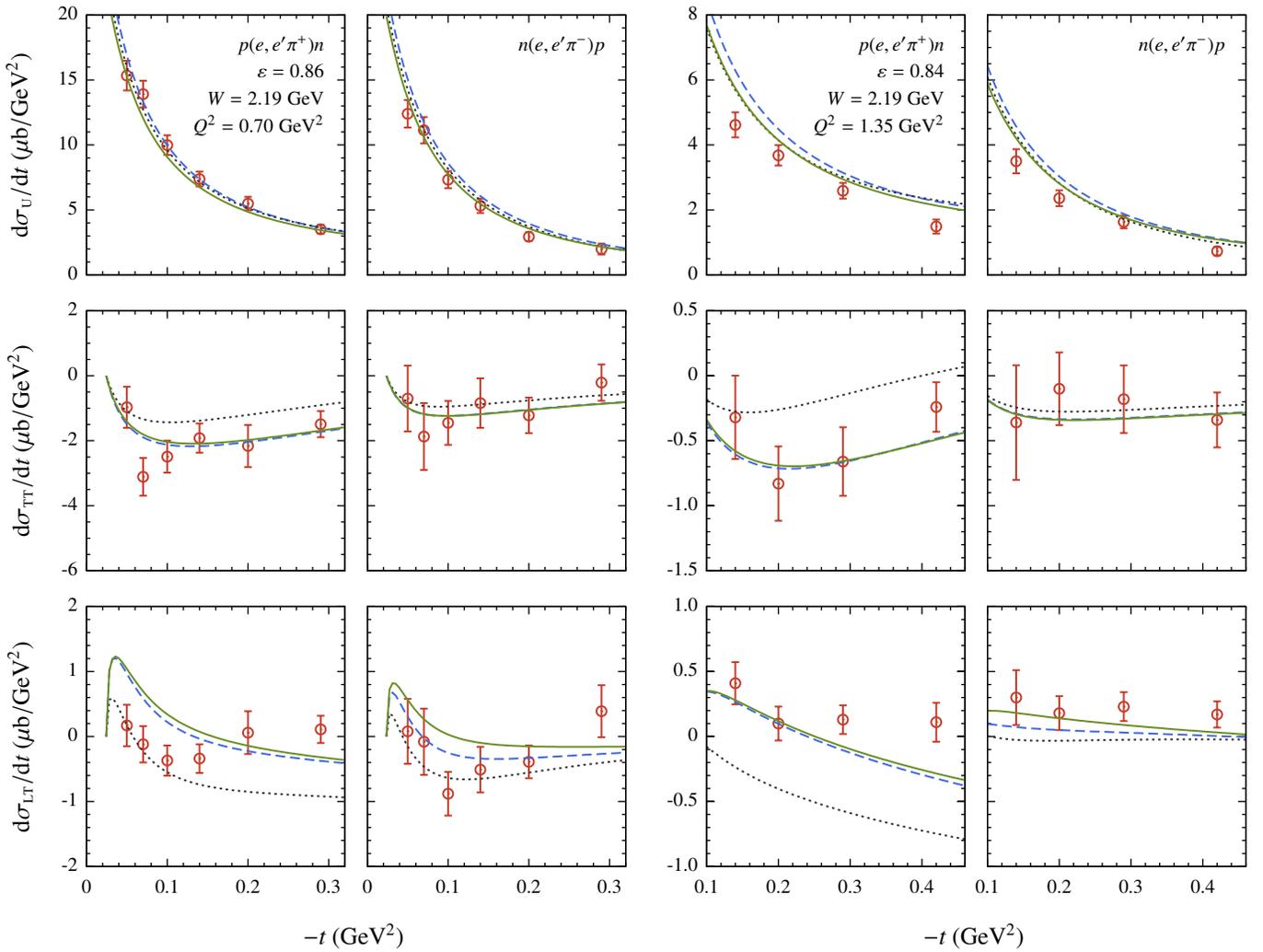

\begin{center}
\footnotesize
\qquad\qquad
\input{DESY-p_diff_U-A-fig}\quad
\input{DESY-n_diff_U-A-fig}\qquad\qquad
\input{DESY-p_diff_U-B-fig}\quad
\input{DESY-n_diff_U-B-fig}
\\ $ $ \\
\qquad\qquad
\input{DESY-p_diff_TT-A-fig}\quad
\input{DESY-n_diff_TT-A-fig}\qquad\qquad
\input{DESY-p_diff_TT-B-fig}\quad
\input{DESY-n_diff_TT-B-fig}
\\ $ $ \\
\qquad\qquad
\input{DESY-p_diff_LT-A-fig}\quad
\input{DESY-n_diff_LT-A-fig}\qquad\qquad
\input{DESY-p_diff_LT-B-fig}\quad
\input{DESY-n_diff_LT-B-fig}
\\ $ $ \\ $ $ \\
\end{center}
\caption{(Color online) The $-t$ dependence of the unseparated and interference cross sections $\d\s{u}/\d t$, and $\d\s{tt}/\d t$ and $\d\s{lt}/\d t$ of the $p(e,e'\pi^+)n$ and $n(e,e'\pi^-)p$ reactions at two different $(W, Q^2, \varepsilon)$ values. Curve notations of Fig.~\ref{fig:CEA-diff_TT-LT} are used. The data are from Ref.~\cite{Brauel:1979zk} (DESY).}
\label{fig:DESY-diff_U-TT-LT}
\end{figure*}

\begin{figure*}[!t]
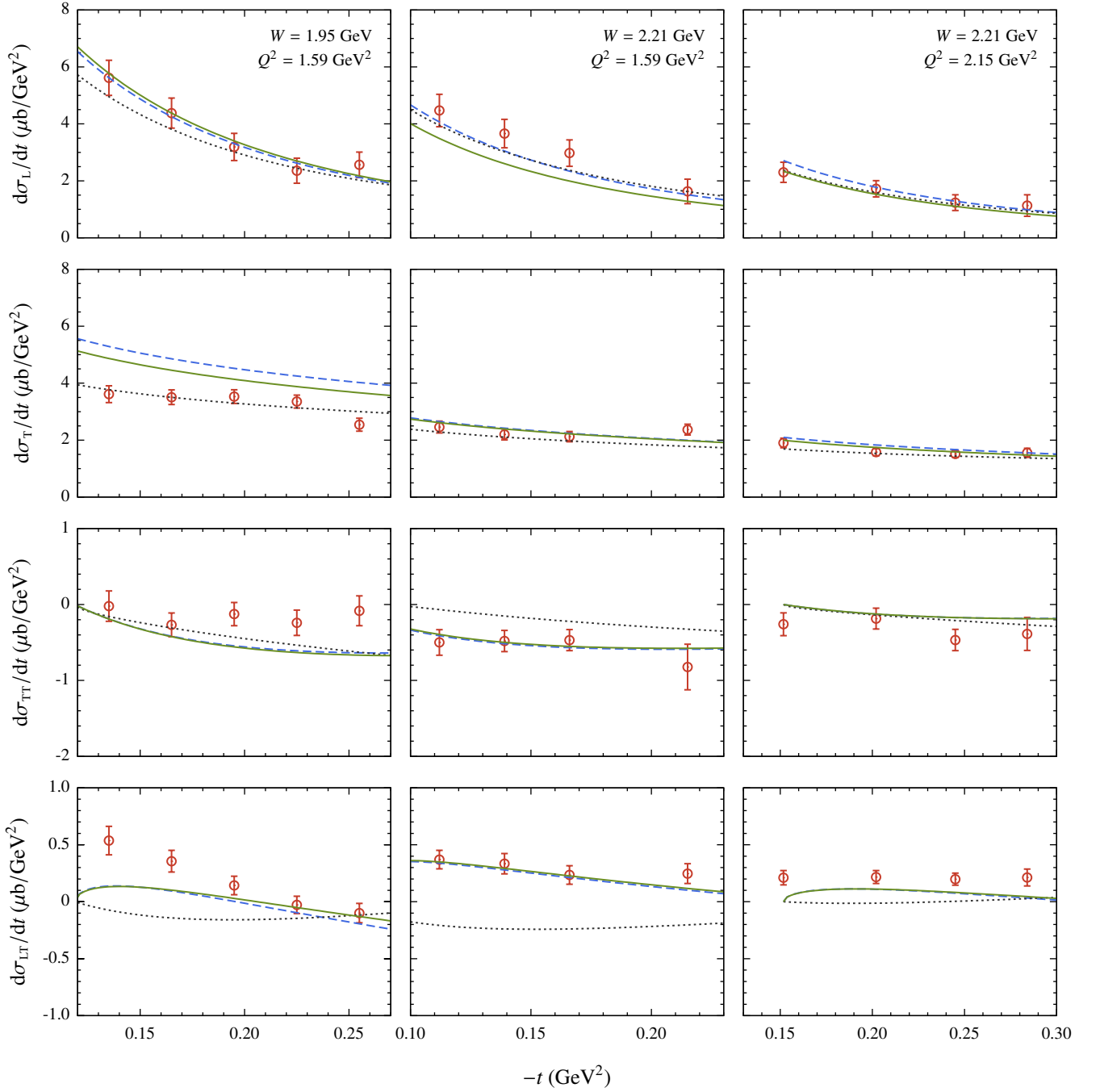

\begin{center}
\footnotesize
\qquad\qquad
\input{JLab-Fpi-1-diff_L-fig}\quad
\input{JLab-Fpi-2-diff_L-fig}\quad
\input{JLab-pi-CT-diff_L-fig}
\\ $ $ \\
\qquad\qquad
\input{JLab-Fpi-1-diff_T-fig}\quad
\input{JLab-Fpi-2-diff_T-fig}\quad
\input{JLab-pi-CT-diff_T-fig}
\\ $ $ \\
\qquad\qquad
\input{JLab-Fpi-1-diff_TT-fig}\quad
\input{JLab-Fpi-2-diff_TT-fig}\quad
\input{JLab-pi-CT-diff_TT-fig}
\\ $ $ \\
\qquad\qquad
\input{JLab-Fpi-1-diff_LT-fig}\quad
\input{JLab-Fpi-2-diff_LT-fig}\quad
\input{JLab-pi-CT-diff_LT-fig}
\\ $ $ \\ $ $ \\
\end{center}
\caption{(Color online) The $-t$ dependence of the longitudinal $\s{l}$, transverse $\s{t}$, and interference cross sections $\s{tt}$ of the $p(e,e'\pi^+)n$ at three different $(W, Q^2)$ values. The $(W, Q^2)$ values listed are the average values; the curves were calculated for the $(W, Q^2)$ values corresponding with the first $-t$ bin. Curve notations of Fig.~\ref{fig:CEA-diff_TT-LT} are used. The data are from the F$\pi$-1 \cite{Tadevosyan:2007yd} (left), F$\pi$-2 \cite{Blok:2008jy} (center), and $\pi$-CT \cite{Horn:2007ug} (right) experiments at JLab.}
\label{fig:JLab-diff_L-T-TT-LT}
\end{figure*}

The VR model can be conceived as a real competitor for the KM model. Despite the fact that it features one additional parameter, the KM model does not provide the best agreement with the data. The VR model, which employs an intuitive prescription for the proton transition form factor, a fixed value for$\Lambda_{\gamma\pi\pi}$, and only one parameter, performs better ($\chi^2_\textsc{ndf} = 3.33$). In Figs.~\ref{fig:CEA-diff_TT-LT} through \ref{fig:JLab-diff_L-T-TT-LT}, the predictions of the KM, the cKM, and the VR model are compared with the experimental DIS data at low $-t$. It is seen that the cKM model offers the worst description of the data and that the predictions of the KM and the VR model are comparably good. The latter provides a slightly better description though.

\begin{figure*}[!t]
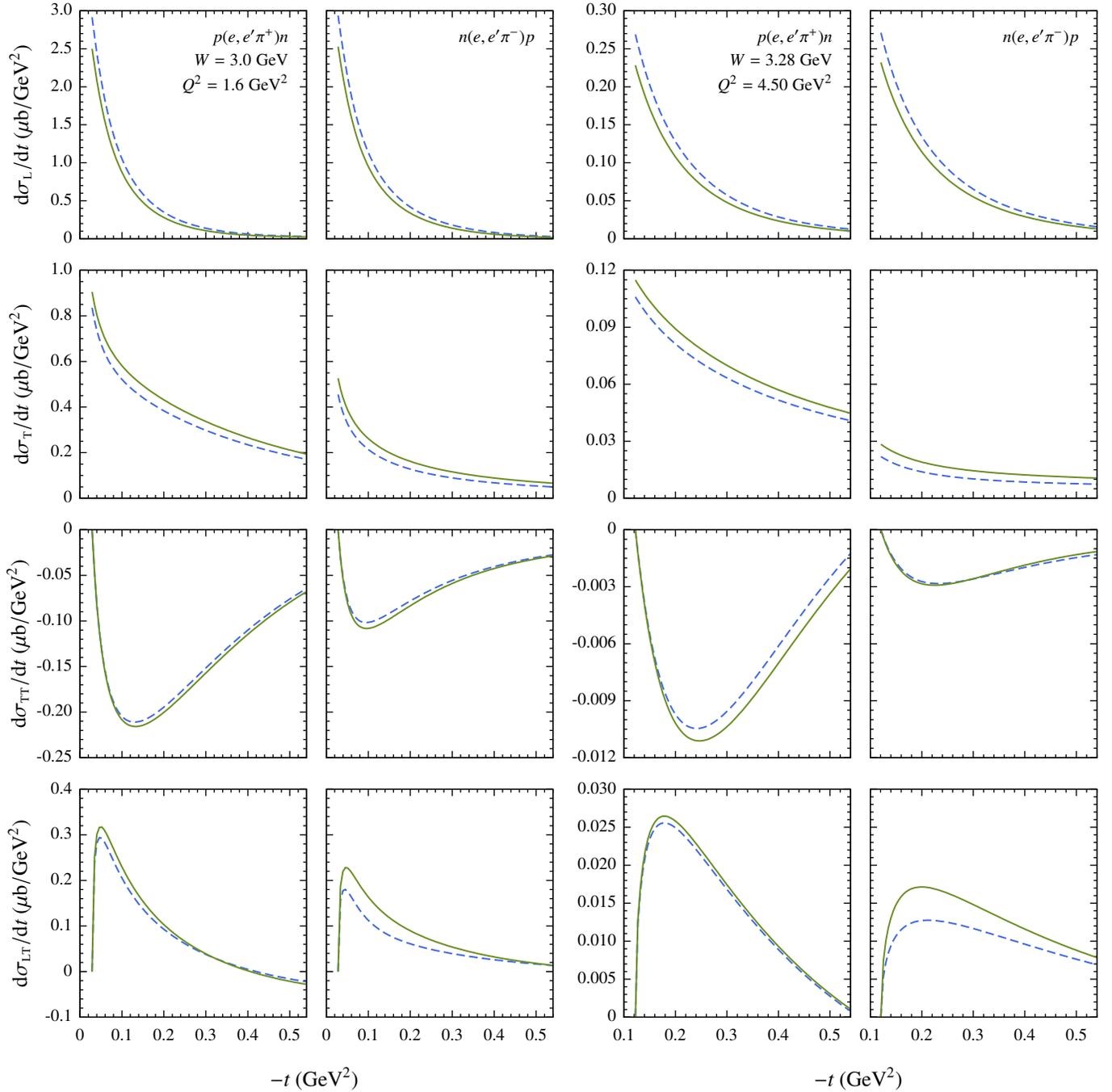

\begin{center}
\footnotesize
\qquad\qquad
\input{JLab-12GeV-p_diff_L-A-fig}\quad
\input{JLab-12GeV-n_diff_L-A-fig}\qquad\qquad
\input{JLab-12GeV-p_diff_L-B-fig}\quad
\input{JLab-12GeV-n_diff_L-B-fig}
\\ $ $ \\
\qquad\qquad
\input{JLab-12GeV-p_diff_T-A-fig}\quad
\input{JLab-12GeV-n_diff_T-A-fig}\qquad\qquad
\input{JLab-12GeV-p_diff_T-B-fig}\quad
\input{JLab-12GeV-n_diff_T-B-fig}
\\ $ $ \\
\qquad\qquad
\input{JLab-12GeV-p_diff_TT-A-fig}\quad
\input{JLab-12GeV-n_diff_TT-A-fig}\qquad\qquad
\input{JLab-12GeV-p_diff_TT-B-fig}\quad
\input{JLab-12GeV-n_diff_TT-B-fig}
\\ $ $ \\
\qquad\qquad
\input{JLab-12GeV-p_diff_LT-A-fig}\quad
\input{JLab-12GeV-n_diff_LT-A-fig}\qquad\qquad
\input{JLab-12GeV-p_diff_LT-B-fig}\quad
\input{JLab-12GeV-n_diff_LT-B-fig}
\\ $ $ \\ $ $ \\
\end{center}
\caption{(Color online) The $-t$ dependence of the separated cross sections $\d\s{l}/\d t$, $\d\s{t}/\d t$, $\d\s{tt}/\d t$, and $\d\s{lt}/\d t$ of the $p(e,e'\pi^+)n$ and $n(e,e'\pi^-)p$ reactions at two different $(W, Q^2)$ values. Curve notations of Fig.~\ref{fig:CEA-diff_TT-LT} are used. These are predictions for the F$\pi$ experiment planned for the 12 GeV upgrade at JLab.}
\label{fig:JLab-12GeV-diff_L-T-TT-LT}
\end{figure*}

In Ref.~\cite{Kaskulov:2010kf}, predictions for the planned F$\pi$ experiment at JLab's 12 GeV upgrade (\cite{Huber:2006pr}) are provided. These are shown for two $(W, Q^2)$ bins in Fig.~\ref{fig:JLab-12GeV-diff_L-T-TT-LT}, together with the corresponding VR predictions. It appears that both models are qualitatively equivalent at these kinematics. There is, however, a quantitative difference between the models, which can become quite substantial in some kinematic regions. For example, at $W = 3.28\GeV$, $Q^2 = 4.50\GeV^2$, and $t = -1.98\GeV^2$ the KM and VR $n(e,e'\pi^-)p$ predictions for $\d\s{lt}/\d t$ differ by about 25\%.

\begin{figure*}[!H]
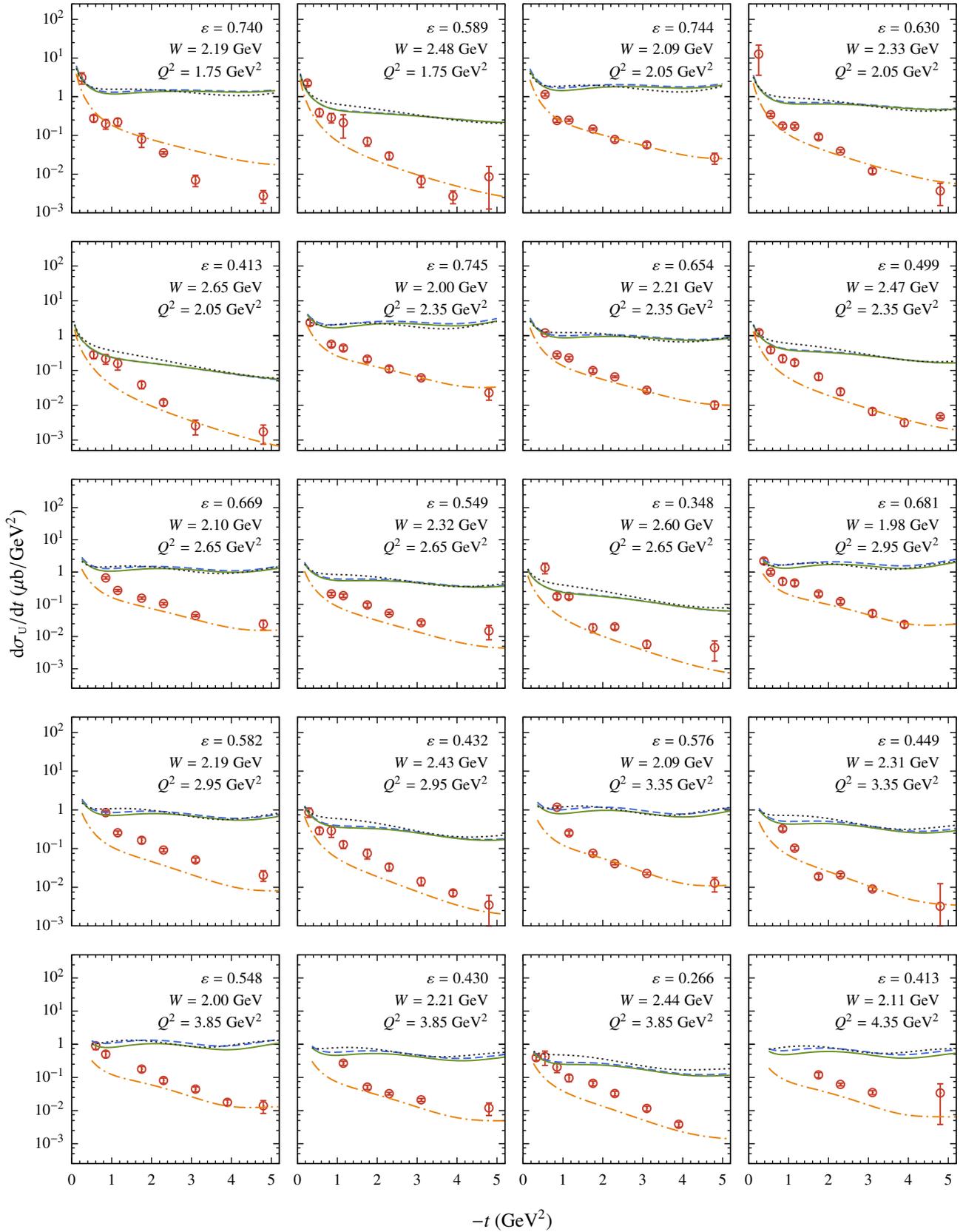

\begin{center}
\footnotesize
\quad
\input{JLab-CLAS-diff_U-A1-fig}\quad
\input{JLab-CLAS-diff_U-B1-fig}\quad
\input{JLab-CLAS-diff_U-C1-fig}\quad
\input{JLab-CLAS-diff_U-D1-fig}
\\ $ $ \\
\quad
\input{JLab-CLAS-diff_U-A2-fig}\quad
\input{JLab-CLAS-diff_U-B2-fig}\quad
\input{JLab-CLAS-diff_U-C2-fig}\quad
\input{JLab-CLAS-diff_U-D2-fig}
\\ $ $ \\
\quad
\input{JLab-CLAS-diff_U-A3-fig}\quad
\input{JLab-CLAS-diff_U-B3-fig}\quad
\input{JLab-CLAS-diff_U-C3-fig}\quad
\input{JLab-CLAS-diff_U-D3-fig}
\\ $ $ \\
\quad
\input{JLab-CLAS-diff_U-A4-fig}\quad
\input{JLab-CLAS-diff_U-B4-fig}\quad
\input{JLab-CLAS-diff_U-C4-fig}\quad
\input{JLab-CLAS-diff_U-D4-fig}
\\ $ $ \\
\quad
\input{JLab-CLAS-diff_U-A5-fig}\quad
\input{JLab-CLAS-diff_U-B5-fig}\quad
\input{JLab-CLAS-diff_U-C5-fig}\quad
\input{JLab-CLAS-diff_U-D5-fig}
\\ $ $ \\ $ $ \\
\end{center}
\caption{(Color online) The $-t$ dependence of the unseparated cross section $\d\s{u}/\d t$ of the $p(e,e'\pi^+)n$ reaction at twenty $(W, Q^2, \varepsilon)$ combinations. The dash-dotted curves correspond with the prediction of the VR model with the inclusion of the strong hadronic form factor $F_{m_t}(t)$ of Eq.~\eq{eq:F_t} with $\Lambda_{m_t} = 0.8 \GeV$; for the other curves the notation of Fig.~\ref{fig:CEA-diff_TT-LT} is used. The data are from Ref.~\cite{Park:2012rn}.}
\label{fig:CLAS_diff_U}
\end{figure*}

\subsection{High \texorpdfstring{$-t$}{-t} regime}
\label{subsec:high_t}
Earlier this year, the CLAS Collaboration at Jefferson Lab published new DIS data for the $p(e,e'\pi^+)n$ reaction \cite{Park:2012rn}. These data cover $-t$ values up to $4.8\GeV^2$ and allow to study the reaction in the deep pion momentum transfer regime. In Fig.~\ref{fig:CLAS_diff_U}, the model predictions are compared with the deep $(Q^2, W, -t)$ CLAS data. It is seen that all three models dramatically overshoot the $-t \gtrsim 1\GeV^2$ data; the corresponding $\chi^2_\textsc{ndf}$ values are $3.9\times 10^4$ (KM), $3.1\times 10^4$ (cKM), and $3.3\times 10^4$ (VR). The data show a much faster fall-off with $-t$, compared to the theoretical curves.

The situation can be remedied to some extent by introducing a form factor in the strong vertex of the $t$-channel Regge amplitudes. Such a strong hadronic form factor accounts for the finite size of the interacting hadrons at the vertex and, in essence, suppresses the $\pi N N$ coupling at high momentum transfers. A possible parametrization for the hadronic form factor is a monopole:
\begin{align}
F_{m_t}(t) = \Biggl(1 + \frac{m_t^2 - t}{\Lambda_{m_t}^2}\Biggr)^{-1},
\label{eq:F_t}
\end{align}
with $m_t \in \{m_\pi, m_\rho, m_{a_1}\}$ the `ground state' mass of the exchanged Regge trajectory and $\Lambda_{m_t}$ the corresponding strong cutoff energy. The finite size of a certain vertex can be accounted for by introducing a running coupling strength. At the $\pi N N$ vertex, for example, this implies that the strong coupling constant $g_{\pi N N}$ acquires a $t$ dependence:
\begin{align}
g_{\pi N N}(t) = g_{\pi N N} F_{m_t}(t). \label{eq:running_coupling}
\end{align}
By construction, one has $F_{m_t}(m_t^2) = 1$. Note that the gauge-fixing $s$-channel term of the pion-exchange current \eq{eq:J_p} is also affected by the strong hadronic form factor, as this term is proportional to $g_{\pi N N}$.

After introducing a coupling of the type \eq{eq:running_coupling} with $\Lambda_{m_t} = 0.8 \GeV$, the VR model provides a reasonable description of the data. Indeed, the dash-dotted curves in Fig.~\ref{fig:CLAS_diff_U} correspond with a $\chi^2_\textsc{ndf} = 16.3$, which is about 2000 times smaller than the $\chi^2_\textsc{ndf}$ value obtained with $F_{m_t}(t) = 1$. The incorporation of the strong hadronic form factor deteriorates the agreement with the DIS data at low $-t$: the $\chi^2_\textsc{ndf}$ value of $3.33$ (see Table \ref{tab:chi2}) is increased to $21.9$ due to the inclusion of $F_{m_t}(t)$.

In Ref.~\cite{Park:2012rn}, it is shown that the predictions of the Laget model \cite{Laget:2004qu} are in fair agreement with the observed unseparated cross sections at high $-t$. Contrary to the KM framework, the Laget model does not consider off-shell effects in the proton electromagnetic transition form factor due to resonances/partons. The Laget model features Reggeized $\pi$ and $\rho$ meson exchanges in the $t$-channel, complemented with the exchange of a nucleon Regge trajectory in the $u$-channel \cite{Laget:2000gj}. The pion cutoff energy $\Lambda_{\gamma\pi\pi}$ is assigned a phenomenological $t$ dependence, which is vital for explaining the observed behavior of $\d\s{u}/\d t$ for $0.5 \lesssim -t \lesssim 5 \GeV^2$ in the Laget model.

In the current framework, the addition of a nucleon Regge trajectory in the $u$-channel and/or the inclusion of a $t$-dependent pion cutoff energy does not considerably improve the description of the high $-t$ data (compared to taking into account a strong hadronic form factor only, which is also included in the Laget model). For now, it is not clear how in the current framework the low $-t$ regime, which does not require a strong hadronic form factor, can be smoothly conjoined with the high $-t$ regime, where $F_{m_t}(t)$ is essential to capture the observed $t$ dependence of $\s{u}$. The corresponding interference cross sections $\s{tt}$ and $\s{lt}$ are the subject of an ongoing analysis by the CLAS Collaboration \cite{Park:2012rn}, and is expected to provide new constraints for the models at high $-t$.

\section{Conclusion and outlook}
\label{sec:conclusions}

In this work, a phenomenological model for the $N(e,e'\pi^\pm)N'$ reaction in the deep-inelastic regime was presented. The model builds on the Kaskulov-Mosel model, which includes three Reggeized background amplitudes in the $t$-channel and takes into account the residual effects of resonances/partons, which are encoded in the proton electromagnetic transition form factor.

It was pointed out that the KM transition form factor, which is derived from duality arguments , does not respect the expected limit for $s \to m_p^2$ by construction. Another suboptimal feature of the KM model is that it uses a discontinuous functional form for the pion cutoff energy. A modified model was proposed, dubbed the ``VR'' model, which resolves both issues. In this model the pion cutoff energy is kept fixed and an intuitive functional dependence for the proton transition form factor was introduced, which respects the physical $s \to m_p^2$ constraints. The VR model has one parameter; the KM model has two. Nevertheless, the VR model offers a somewhat better description of the low $-t$ $N(e,e'\pi^\pm)N'$ data than the KM one. It was shown that imposing the correct $s \to m_p^2$ behavior in the KM prescription for the proton electromagnetic transition form factor, does not result in a fair description of the data.

The VR predictions for the planned F$\pi$ experiment at JLab were provided and compared to those of the KM model. The models were also tested against the recent unseparated $p(e,e'\pi^+)n$ data which extend to pion momentum transfers of $4.8\GeV^2$, corresponding with the deep $-t$ regime. The VR and KM models fail miserably to describe the observed $t$ dependence of the cross sections. For the VR model, it was shown that the introduction of a strong hadronic form factor in the $\pi N N$ vertex dramatically improves the agreement with the high $-t$ data, but at the same time results in a deteriorated description of the low $-t$ data. At this moment it is unclear how the low and the high $-t$ regime can be smoothly matched. It can be expected that the availability of separated cross sections at high $-t$ will shed new light on this issue.


\acknowledgments
This work is supported by the Research Council of Ghent University and the Flemish Research Foundation (FWO Vlaanderen). The authors would like to thank Dipangar Dutta, Tanja Horn, Garth Huber, and Igor Strakovsky for providing most of the relevant data, and Murat Kaskulov for his help with the implementation of the KM model.


\end{document}